\documentclass[pss]{wiley2sp} % provides new 2008 pss two-column style (no alternative manuscript style output available at present)
\usepackage{amssymb,amsmath} 	 
\usepackage[T1]{fontenc}
\usepackage[latin1]{inputenc}
\usepackage[thinspace,thinqspace,squaren]{SIunits}					%nice units and spacing

\makeatletter 
\DeclareRobustCommand*\textsubscript[1]{%
  \@textsubscript{\selectfont#1}}
\newcommand{\@textsubscript}[1]{%
  {\m@th\ensuremath{_{\mbox{\fontsize\sf@size\z@#1}}}}}
\makeatother
\newcommand*{\Molek}[2]{{#1\textsubscript{{#2}}}}										%Molek
\newcommand{\LRS}{\Molek{Lu}{}\Molek{Rh}{2}\Molek{Si}{2}}							%LRS
\newcommand{\YRS}{\Molek{Yb}{}\Molek{Rh}{2}\Molek{Si}{2}}							%YRS
\newcommand{\RH}{\ensuremath{R_{\mathrm H}}}												%R_H
\newcommand{\rhoH}{\ensuremath{\rho_{\mathrm H}}}										%rho_H
\newcommand{\dd}{\mathrm{d}} 																	%Differential d
\newcommand{\HEunit}[1]{\unit{#1 \cdot 10^{-10} } \cubicmetre\per\coulomb}

\begin{document}

% Title of the article
\title{The crossed-field and single-field 
Hall effect in \LRS}

% Abbreviated title for the page headers
\titlerunning{Crossed-field Hall effect in \LRS}

% Authors
\author{%
  Sven Friedemann\textsuperscript{\Ast,\textsf{\bfseries 1}},
  Niels Oeschler\textsuperscript{\textsf{\bfseries 1}},
  Steffen Wirth\textsuperscript{\textsf{\bfseries 1}},
  Frank Steglich\textsuperscript{\textsf{\bfseries 1}},
  Sam MaQuilon\textsuperscript{\textsf{\bfseries 2}},
  Zachary Fisk\textsuperscript{\textsf{\bfseries 2}}  
  }

% Abbreviated list of authors for the page headers
\authorrunning{Sven Friedemann et al.}

%E-mail-address of corresponding author
\mail{e-mail
  \textsf{Sven.Friedemann@cpfs.mpg.de}, Phone:
  +49-351-46463219, Fax: +49-351-46463232}

% author's affiliations/addresses
\institute{%
  \textsuperscript{1}\,Max Planck Institute for Chemical Physics of Solids, N{\"o}thnitzer Stra{\ss}e 40, 01187 Dresden, Germany\\
  \textsuperscript{2}\,Department of Physics and Astronomy, University of California, Irvine, CA 92697-4575, USA\\
  }

\received{XXXX, revised XXXX, accepted XXXX} % do not change, will be filled in by the publisher
\published{XXXX} % do not change, will be filled in by the publisher

%Please select four to six PACS-codes from the enclosed list (PACS.txt) or from www.aip.org/pacs)
 % For example: 71.20.Ps
\abstract{%
% Usage: \abstcol{<left column>}{<right column>}
\abstcol{%left column
The Hall effect of \LRS---the non-magnetic homologue of the heavy-fermion material \YRS---is studied with two different setups: In the conventional single-field geometry, the field dependence is analyzed in terms of the differential Hall coefficient. 
Beyond that, the recently developed crossed-field experiment allows to examine the linear-response Hall coefficient as a function of magnetic field.
%}{
The results reveal the expected analogy between both experiments which corroborates the equivalent findings in  \YRS. This emphasizes the applicability to investigate field-induced quantum critical points with both methods.}
 }
\titlefigure[width=.45\textwidth]{probenskizze}
\titlefigurecaption{Setup of the crossed-field Hall-effect experiment.}

\maketitle   % please do not remove

\section{Introduction}
\begin{figure*}[htb]
	\setlength{\unitlength}{\textwidth}
	\begin{picture}(.49,.41)
  		\put(0,0){\includegraphics[width=.49\unitlength]{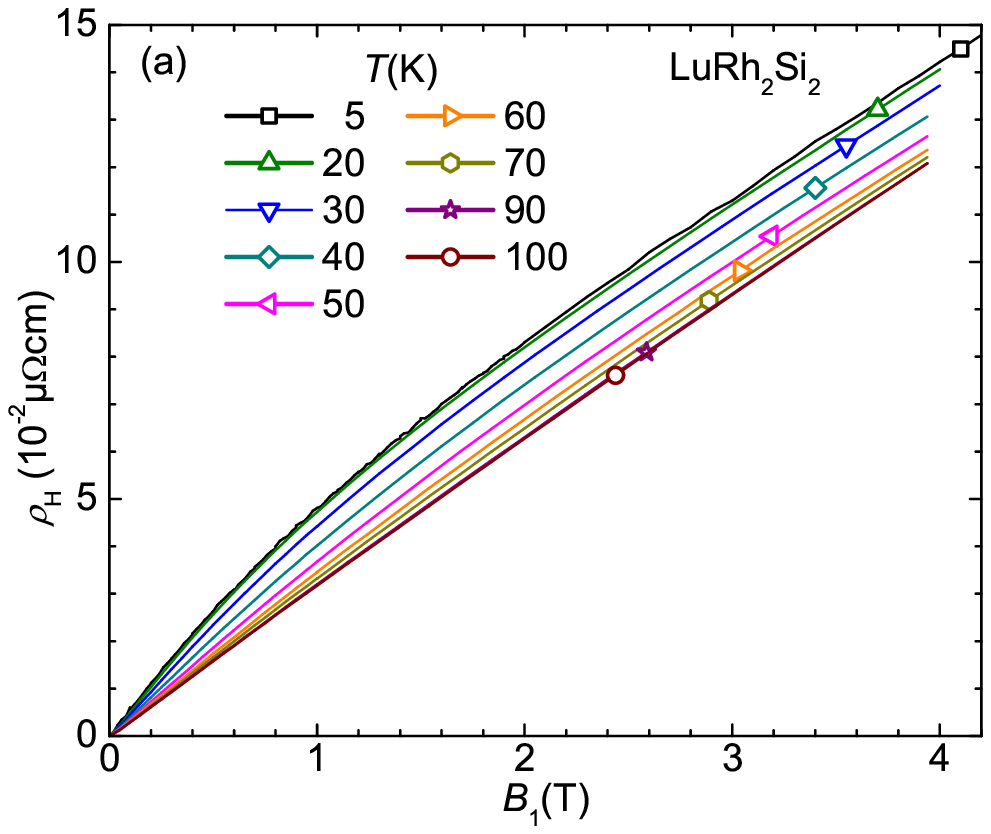}}
  		\put(.185,.06){\includegraphics[width=.3\unitlength]{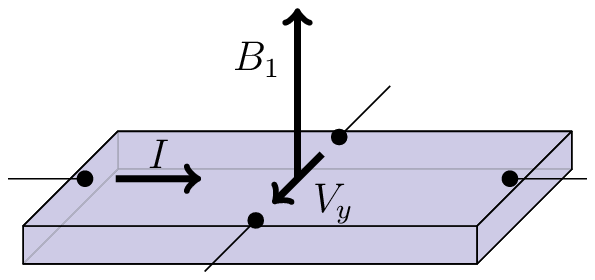}}
  	\end{picture}
  	\hfill
  	\includegraphics[width=.49\unitlength]{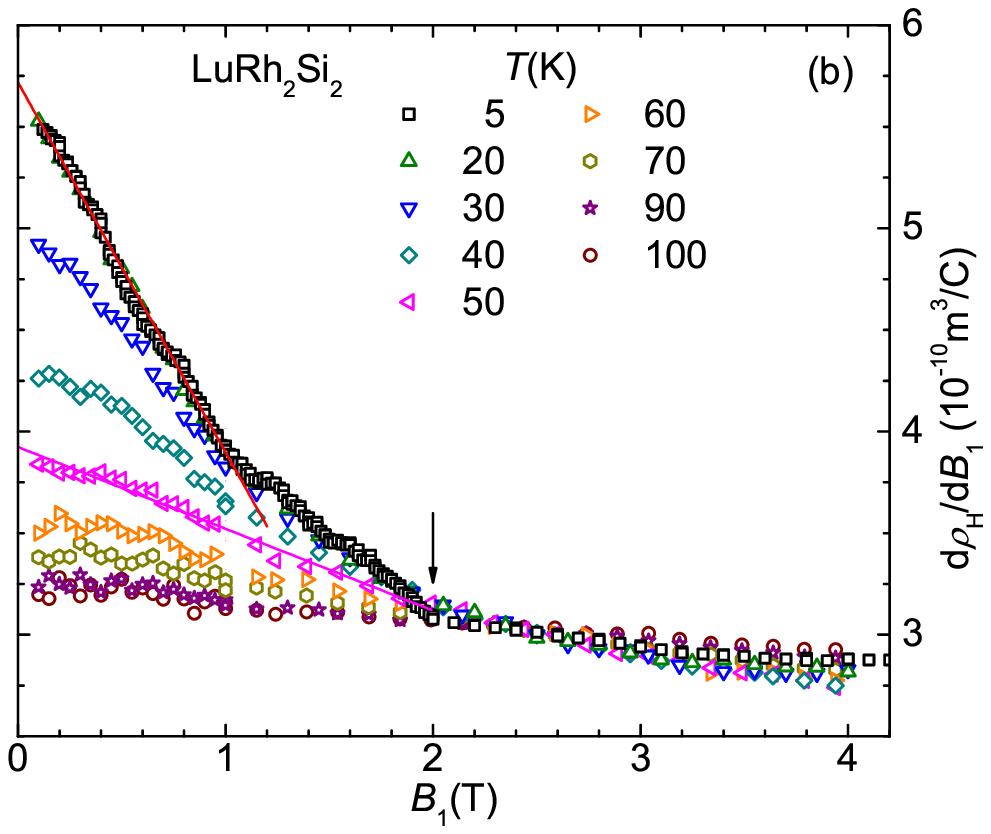}
	\caption{Hall resistivity (a) and differential Hall coefficient (b) of \LRS\ at selected temperatures. Inset sketches the single-field setup. Solid straight lines in (b) are linear fits to the data at \unit{5}\kelvin\ and \unit{50}\kelvin\ for fields below \unit{1}\tesla. Arrow indicates the crossover field.}
	\label{fig:RhoH}
\end{figure*}
Quantum critical points (QCPs) have recently attracted considerable interest \cite{Focus2008}. 
They may arise via the suppression of a continuous phase transition to zero temperature by a non-thermal control parameter. 
For the proper description of QCPs in heavy fermion systems, two scenarios are %currently 
available \cite{Gegenwart2008} predicting either a smooth (the spin-density-wave scenario) or a discontinuous (the Kondo-breakdown scenario) evolution of the Fermi surface at the QCP. 
%Therefore, an experimental study of the Fermi surface is called to distinguish between the two alternatives \cite{Coleman2001}. The Fermi surface can be investigated 
Therefore, the experimental investigation of the Fermi surface is called for to identify the nature of a particular QCP \cite{Coleman2001}. This might be realized
by using de Haas-van Alphen (dHvA) effect, photoemission spectroscopy  or Hall effect. Currently, it is not possible to perform photoemmission spectroscopy measurements at the low temperatures needed to track quantum criticality. The dHvA effect was already used to show a change of the Fermi surface in \Molek{CeRhIn}5 \cite{Shishido2005}. However, dHvA studies are usually limited to high fields, which might in some interesting cases be beyond the QCP. Therefore, the Hall effect is presently the best-suited probe to investigate the Fermi-surface evolution at the QCP \cite{Paschen2006}. Particularly, \YRS\ has emerged as a prototypical material in which the control parameter---in this case the magnetic field---can be tuned continuously \cite{Custers2003}. Moreover, a crossover of the Hall coefficient was observed which sharpens to an abrupt change at the QCP in the zero temperature limit \cite{Paschen2004}. This was considered evidence for the Kondo-breakdown scenario. In these experiments the magnetic field has a dual role: On the one hand, it generates the Hall response from which the Hall coefficient is extracted while on the other hand it tunes the ground state of the material through the QCP. This allows for two different setups: In the conventional, single-field experiment one magnetic field simultaneously performs both tasks (see inset of figure \ref{fig:RhoH}(a)). To extract the non-linear contributions of the Hall effect arising from the tuning through the QCP one has to examine the differential Hall coefficient. However, this separation is not straightforward as an additional term arises due to the Zeeman effect \cite{Paschen2004}. Therefore, the crossed-field setup was designed in which two perpendicular fields are utilized to separate the two effects: One field perpendicular to the current generates the Hall response, whereas the second field is oriented parallel to the current such that it only tunes the ground state (cf.\ inset of figure \ref{fig:RhoHvsB1_LRS_CF}(a)).

Here, we report Hall-effect measurements on \LRS\ ---the non-magnetic homologue to \YRS---providing a crosscheck for the applicability of the two Hall effect setups. 
 
\section{Experimental Setup}
Single crystals of \LRS\ were grown in an indium flux. 
Excess flux was removed with hydrochloric acid. The resistivity was measured to ensure In-free samples. 
The samples were polished to thin platelets with thickness $t < \unit{60}\micro\metre$. The current $I$ flowed within the crystallographic $ab$ plane. The magnetic field $B_1$ was applied along the $c$ axis, generating the Hall voltage within the $ab$ plane. This setup is henceforth called single-field setup and it is sketched in the inset of figure \ref{fig:RhoH}(a).
The Hall resistivity \rhoH\ was calculated as the antisymmetric part of the transverse voltage $V_y$: $\rhoH=t/2I *(V_y(B_1)-V_y(-B_1))$ to correct for misalignments of the contacts. 
Subsequently, the differential Hall coefficient was numerically calculated as $\dd \rhoH (B_1)/\dd B_1$ for the single-field experiments. Single-field measurements were performed in a Quantum Design Physical Property Measurements System down to \unit{5}\kelvin.

For the so-called crossed-field experiments an additional magnetic field $B_2$ parallel to the current was used
(see inset of figure \ref{fig:RhoHvsB1_LRS_CF}(a) for a sketch of the setup). 
This field cannot induce a Hall response as it is parallel to the current.
In the case of the crossed-field experiment, the linear-response
Hall coefficient $\RH(B_2)$ was derived as the slope of linear fits to the Hall resistivity $\rhoH(B_1)|_{B_2}$ at fixed field $B_2$ for fields $B_1\leq \unit{0.4}\tesla$ as illustrated in figure \ref{fig:RhoHvsB1_LRS_CF}(a). Crossed-field experiments were conducted in a $^3$He/\,$^4$He-dilution refrigerator at temperatures $T \geq \unit{20}\milli\kelvin$.

In both, the single-field and the crossed-field setup, the Hall response is generated by the magnetic field $B_1$. A tuning effect to the ground state of the sample on the other hand may arise from both $B_1$ and $B_2$. A separation of the two effects may be achieved in the crossed-field setup if the tuning effect of $B_1$ is much smaller than that of $B_2$. In the case of \YRS, the strong magnetic anisotropy was utilized to ensure this condition. For the here presented measurements on \LRS\ this condition will have to be surmised.

\section{Results and Discussion}
\subsection{Single-field}
The Hall resistivity \rhoH\ measured in single-field setup and the thereof derived differential Hall coefficient $\dd\rhoH/\dd B_1$ are presented in figure \ref{fig:RhoH}(a) and (b), respectively. Three temperature regimes are observed: In both the low temperature range below \unit{20}\kelvin\ and in the high-temperature range above \unit{70}\kelvin, no temperature dependence of the Hall resistivity is observed. Only at intermediate temperatures a crossover between the characteristics of the two limiting cases is seen as will be discussed below.

\begin{figure*}%
   \setlength{\unitlength}{.98\textwidth}
	\begin{picture}(.5,.417)
  		\put(0,0){\includegraphics[width=.5\unitlength]{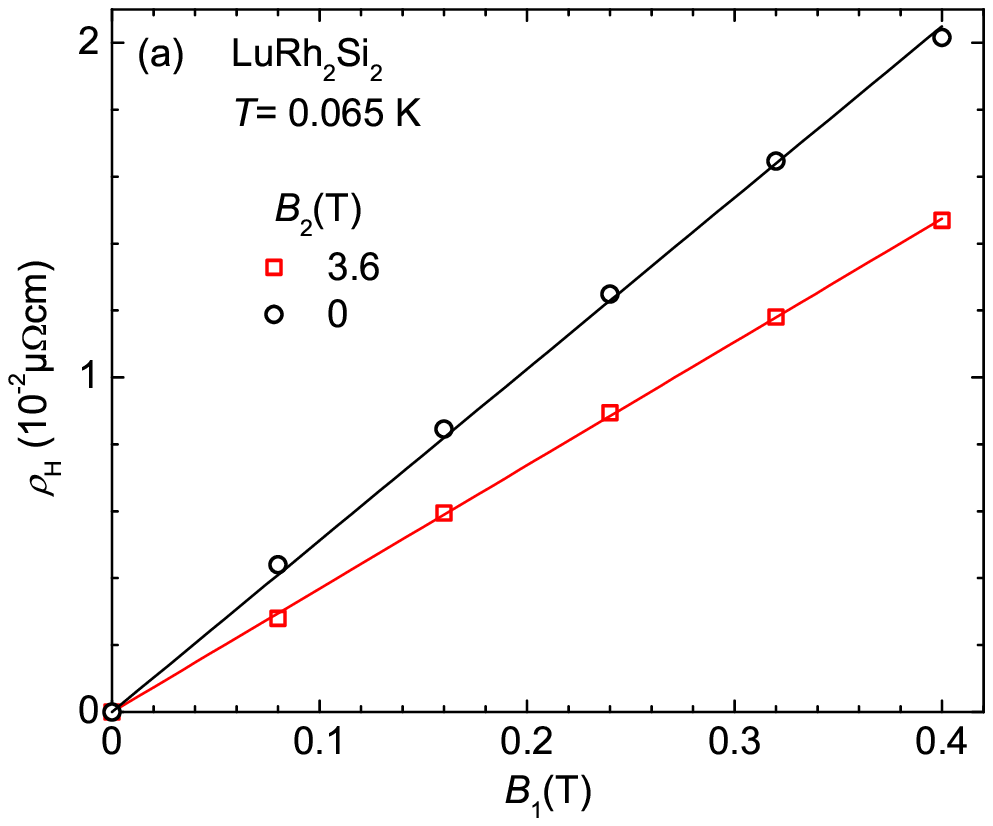}}
  		\put(.195,.06){\includegraphics[width=.3\unitlength]{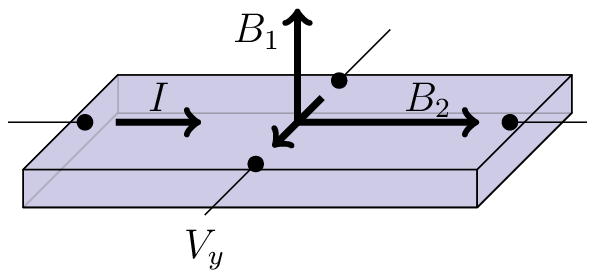}}
  \end{picture}
	\hfill
	\includegraphics[width=.5\unitlength]{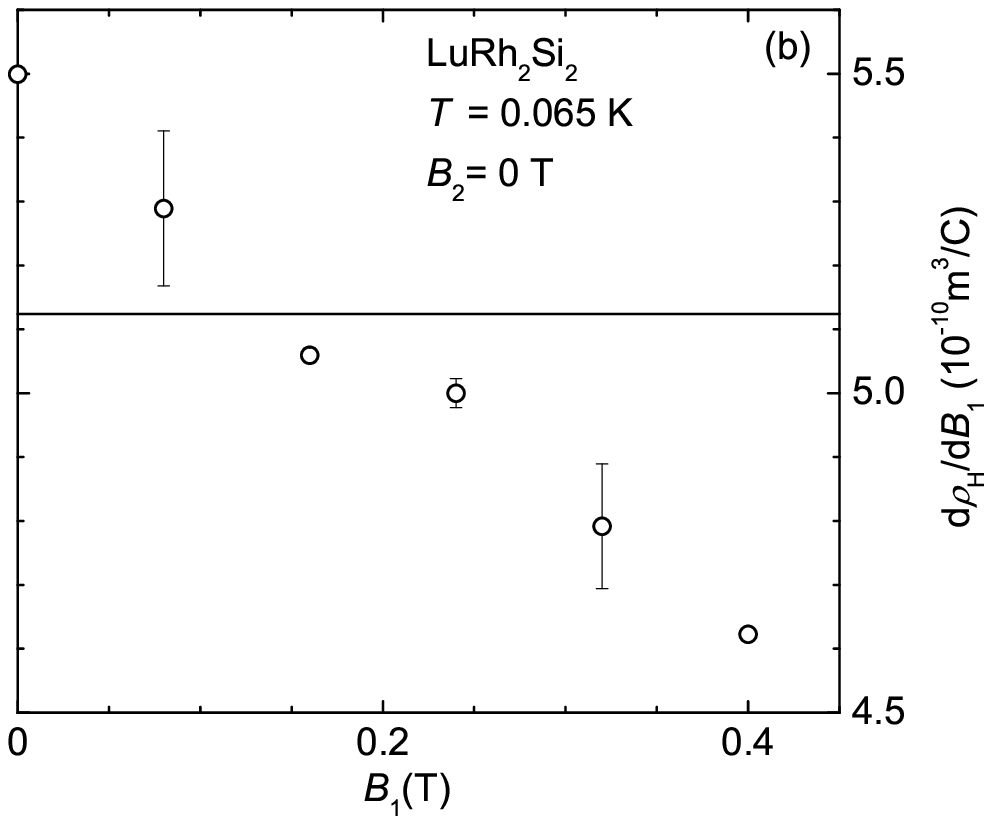}%
	\caption{(a) Hall resistivity $\rhoH(B_1)|_{B_2}$ at a fixed temperature of \unit{0.065}\kelvin\ for two selected fields $B_2$. Lines are linear fits to the data from which the Hall coefficient as shown in figure \ref{fig:RHvsB2} was calculated. Inset sketches the crossed-field setup.
	Differential Hall coefficient of the $B_2=\unit{0}\tesla$ curve of panel (a). Solid line corresponds to linear fit of panel (a) and directly reflects the Hall coefficient extracted (cf. figure \ref{fig:RHvsB2}).}%
	\label{fig:RhoHvsB1_LRS_CF}%
\end{figure*}

Within the low-temperature range, the field dependence of the Hall resistivity exhibits two regimes with different slopes: A large initial slope is observed below \unit{1}\tesla, followed by a crossover to linearity for fields above \unit{2}\tesla. In the intermediate temperature range, \textit{i.e.}\ for $\unit{20}\kelvin \leq T \leq \unit{70}\kelvin$, the different field regimes evolve differently: The high-field slope remains unchanged throughout the whole temperature range investigated. By contrast, the initial slope is found to decrease as the temperature is increased. Upon approaching the high-temperature range, the low-field slope merges with that of the high-field regime. As a consequence, the crossover is absent at high temperatures, $T\geq\unit{70}\kelvin$.

The two field regimes are more obvious in the differential Hall coefficient in figure \ref{fig:RhoH}(b): $\dd\rhoH/\dd B_1$ follows a linear decrease with increasing field $B_1$ for fields below \unit{1}\tesla\ with the highest slope observed in the low temperature range, $T\leq \unit{20}\kelvin$. At fields above \unit{2}\tesla, the differential Hall coefficient is almost constant at \HEunit{3}. In the intermediate temperature range, the low-field values and accordingly the slope of the linear field dependence are found to decrease with increasing temperature. As the high-temperature range is approached, the differential Hall coefficient $\dd \rhoH / \dd B_1$ converges to the high-field value. Consequently, the differential Hall coefficient depicts a unique field dependence without a crossover for temperatures above \unit{70}\kelvin. The decrease of the differential Hall coefficient at small fields reflects the decrease of the low-field Hall coefficient as a function of temperature between \unit{20}\kelvin\ and \unit{70}\kelvin\ \cite{Friedemann2008}. 

The crossover field does not change as a function of temperature as clearly revealed by the differential Hall coefficient (figure \ref{fig:RhoH}(b)). This allows to rule out a low-field to high-field transition of the Hall effect expected at $\omega_{\mathrm c}\tau \approx 1$ with the cyclotron frequency $\omega_{\mathrm c} \propto B_1$ of the electron orbits and the scattering time $\tau \propto \rho^{-1}$ \cite{Hurd1972}. The transition field $B_1$ where the condition $\omega_{\mathrm c}\tau \propto B_1 / \rho \approx 1$ is fulfilled should increase with increasing resistivity $\rho$. The resistivity of \LRS\ rises by a factor of 5 as the temperature is raised from \unit{10}\kelvin\ to \unit{100}\kelvin. Consequently, the robustness of the observed crossover is incompatible with a low-field to high-field transition. This is supported by an estimate of the transition field via $\omega_{\mathrm c}\tau = B_1/(n e \rho_0) \approx 1$ with the residual resistivity $\rho_0= \unit{1.2}\micro\ohm\usk\centi\metre$, the elementary charge $e$ and the carrier density $n$ from band structure calculations \cite{Friedemann2008a} yielding a transition field exceeding \unit{70}\tesla, which is far beyond the accessed field range. One might speculate that the crossover is rather caused by the two-band behavior identified in the temperature dependence of \LRS\ \cite{Friedemann2008a} as indicated by the fact that the high-field slope matches the initial slope at high temperatures. However, it remains to be understood why a field of \unit{2}\tesla\ has a comparable effect like warming up to \unit{100}\kelvin.

\subsection{Crossed-field}
\begin{figure*}
  \sidecaption
  	\includegraphics[width=.59\textwidth]{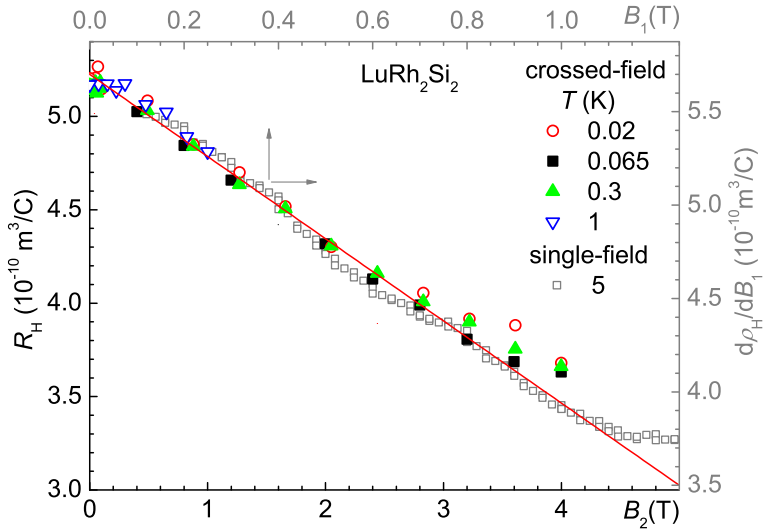}
   \caption{Crossed-field Hall coefficient of \LRS\ as a function of the magnetic field $B_2$ at selected temperatures (left and bottom axes). The solid line represents a linear fit to all datasets for fields below \unit{3}\tesla.  Single-field data at \unit{5}\kelvin\ from figure \ref{fig:RhoH}(a) are replotted in gray (right and top scale) with the field axis scaled by $1/4$ and the ordinate shifted by \HEunit{0.475} accounting for the reduction of \RH\ in the crossed-field experiment (see text).}
  \label{fig:RHvsB2}
\end{figure*}
The tuning-field dependence of the Hall coefficient, $\RH(B_2)$, in the crossed-field setup is shown in figure \ref{fig:RHvsB2}. In the considered temperature range, \textit{i.e.}\ below \unit{1}\kelvin, all curves are nearly identical which is a continuation of the temperature independence of the differential Hall coefficient observed in the low temperature range below \unit{20}\kelvin\ (figure \ref{fig:RhoH}(b)). At zero field, a value of \HEunit{5.2}\ is measured while \RH\ decreases linearly with increasing $B_2$ for fields up to \unit{3}\tesla\ (solid line in figure \ref{fig:RHvsB2}(a)). At higher fields slight deviations of $\RH(B_2)$ towards higher values compared to the low-field linearity are found.

The comparison of the crossed-field results with those of the single-field experiment in figure \ref{fig:RHvsB2} reveals that both obey a linear decrease at small fields in the low-temperature range. However, the scaling of the field axis of the single-field differential Hall coefficient by a factor of 4 reveals that the crossed-field Hall coefficient exhibits a 4 times smaller slope. Furthermore, the linearity is provided over an almost 4 times larger field range in the crossed-field experiment. This indicates a magnetic anisotropy requiring 4 times larger fields applied within the $ab$ plane (crossed-field experiment) to obtain the same effect like for fields along the $c$ axis (single-field experiment). This anisotropy would naturally explain both the decreased slope and an enlarged field range fulfilling the linearity. The deviations of $\RH(B_2)$ from linearity above \unit{3}\tesla\ reflect the corresponding behavior of the single-field results at \unit{1}\tesla. The crossover towards constant \RH\ is expected at fields beyond the accessible range of our crossed-field measurements. 

We note that the above discussed magnetic anisotropy of \LRS\ is opposite to that of \YRS. Consequently, the tuning effect of $B_1$ is not negligible and leads to the slight deviations from linearity observed for the Hall resistivity $\rhoH(B_1)$ presented in figure \ref{fig:RhoHvsB1_LRS_CF}(a). This is more obvious in figure \ref{fig:RhoHvsB1_LRS_CF}(b) as a decreasing differential Hall coefficient upon increasing field $B_1$. The low-field value is then in good agreement with the results of the single-field experiment (figure \ref{fig:RhoH}(b)). Furthermore, figure \ref{fig:RhoHvsB1_LRS_CF}(b) illustrates the systematic deviations of the linear-slope Hall coefficient extracted in the crossed-field measurements compared to the differential Hall coefficient extracted in the single-field experiment: In the crossed-field case, the Hall coefficient was determined as the slope of the linear fits for fields $B_1$ up to \unit{0.4}\tesla\ (horizontal line in figure \ref{fig:RhoHvsB1_LRS_CF}(b)) which represents an average over the differential Hall coefficient in this field range (data in figure \ref{fig:RhoHvsB1_LRS_CF}(b)). The decrease of the differential Hall coefficient with increasing field, therefore, causes a reduction of the linear-slope Hall coefficient (crossed-field experiment)  (figure \ref{fig:RHvsB2}(a)) with respect to the differential Hall coefficient (figure \ref{fig:RhoH}(b)). This reduction is accounted for by a shift of the ordinate in the direct comparison of both experiments in figure \ref{fig:RHvsB2}.

\section{Conclusions}
In summary, we emphasize that the crossed-field and the single-field experiment yield consistent results: The field dependences of the Hall coefficient in both experiments is found to be in good agreement if the field axis of the single-field experiment is scaled by a factor of $1/4$. This scaling arises due to the fact that the tuning-fields are applied along different crystallographic directions for the different experiments strongly indicating a magnetic anisotropy of \LRS. The anisotropy is likely caused by the actual band structure which displays a pronounced anisotropy as expected for a tetragonal system. Our comprehensive study establishes both the single-field and the crossed-field experiment as a powerful tool to investigate field induced QCPs. In particular, it would be interesting to study an example for the spin-density-wave scenario which might be realized at the field-induced QCP in CeCu$_{5.8}$Au$_{0.2}$ \cite{Stockert2007}.

\begin{acknowledgement}
The authors would like to thank P. Gegenwart, S. Paschen, T. Westerkamp, G. Wigger and G. Zwicknagl for fruitful discussions. This work was partially supported by NSF-DMR-0710492. Part of the work at Dresden was supported by the 
DFG Research Unit 960 ``Quantum Phase Transitions''.
\end{acknowledgement}

%\bibliography{../../../../Literatur/svenbib}

\begin{thebibliography}{[10]}

\bibitem{Focus2008}% article
 \textsc{Focus}\iffalse Quantum phase transitions\fi,
 \jr{Nature Phys.} \textbf{4}(3), 167--204 (2008).


\bibitem{Gegenwart2008}% article
 \textsc{P.~Gegenwart} \etal{}\iffalse Quantum criticality in heavy-fermion
  metals\fi,
 \jr{Nature Phys.} \textbf{4}(3), 186--197 (2008).


\bibitem{Coleman2001}% article
 \textsc{P.~Coleman} \etal{}\iffalse How do fermi liquids get heavy and
  die?\fi,
 \jr{J. Phys.: Condens. Matter} \textbf{13}(35), R723--R738 (2001).


\bibitem{Shishido2005}% article
 \textsc{H.~Shishido} \etal{}\iffalse A drastic change of the fermi surface at
  a critical pressure in cerhin$_5$: dhva study under pressure\fi,
 \jr{J. Phys. Soc. Jpn.} \textbf{74}(4), 1103--1106 (2005).


\bibitem{Paschen2006}% article
 \textsc{S.~Paschen}\iffalse Hall effect for classification of quantum critical
  points\fi,
 \jr{Physica B} \textbf{378-380}(May), 28--30 (2006).


\bibitem{Custers2003}% article
 \textsc{J.~Custers} \etal{}\iffalse The break-up of heavy electrons at a
  quantum critical point\fi,
 \jr{Nature} \textbf{424}(6948), 524--527 (2003).


\bibitem{Paschen2004}% article
 \textsc{S.~Paschen} \etal{}\iffalse Hall-effect evolution across a
  heavy-fermion quantum critical point\fi,
 \jr{Nature} \textbf{432}(7019), 881--885 (2004).


\bibitem{Friedemann2008}% article
 \textsc{S.~Friedemann} \etal{}\iffalse Band-structure and anomalous
  contributions to the hall effect of ybrh$_2$si$_2$\fi,
 \jr{Physica B} \textbf{403}(5-9), 1251--1253 (2008).


\othercit
\bibitem{Hurd1972}% book
 \textsc{C.\,M. Hurd},
The Hall effect in metals and alloys (Plenum Press, New York, 1972).


\bibitem{Friedemann2008a}% article
 \textsc{S.~Friedemann} \etal{}\iffalse Hall effect measurements on
  ybrh$_2$si$_2$ in the light of electronic structure calculations\fi,
 \jr{arXiv:0803.4428v1} (2008).


\bibitem{Stockert2007}% article
 \textsc{O.~Stockert} \etal{}\iffalse Magnetic fluctuations at a field-induced
  quantum phase transition\fi,
 \jr{Phys. Rev. Lett.} \textbf{99}(23), 237203--4 (2007).


\end{thebibliography}
%\bibliographystyle{pss-mod} 

\providecommand{\WileyBibTextsc}{}
\let\textsc\WileyBibTextsc
\providecommand{\othercit}{}
\providecommand{\jr}[1]{#1}
\providecommand{\etal}{~et~al.}

\end{document}